# The evolutionary origins of modularity

Jeff Clune, Jean-Baptiste Mouret and Hod Lipson

| | |
|---|---|
| **Supplementary data** | "Data Supplement"<br>http://rspb.royalsocietypublishing.org/content/suppl/2013/01/29/rspb.2012.2863.DC1.html |
| **References** | This article cites 34 articles, 9 of which can be accessed free<br>http://rspb.royalsocietypublishing.org/content/280/1755/20122863.full.html#ref-list-1 |
|  EXiS Open Choice | This article is free to access |
| **Subject collections** | Articles on similar topics can be found in the following collections<br><br>computational biology (13 articles)<br>evolution (1383 articles)<br>systems biology (32 articles) |
| **Email alerting service** | Receive free email alerts when new articles cite this article - sign up in the box at the top right-hand corner of the article or click **here** |

To subscribe to *Proc. R. Soc. B* go to: **http://rspb.royalsocietypublishing.org/subscriptions**

**Author for correspondence:**
Jeff Clune
e-mail: jclune@uwyo.edu

# The evolutionary origins of modularity

Jeff Clune[1,2,†], Jean-Baptiste Mouret[3,†] and Hod Lipson[1]

[1]Cornell University, Ithaca, NY, USA
[2]University of Wyoming, Laramie, WY, USA
[3]ISIR, Université Pierre et Marie Curie-Paris 6, CNRS UMR 7222, Paris, France

A central biological question is how natural organisms are so evolvable (capable of quickly adapting to new environments). A key driver of evolvability is the widespread modularity of biological networks—their organization as functional, sparsely connected subunits—but there is no consensus regarding why modularity itself evolved. Although most hypotheses assume indirect selection for evolvability, here we demonstrate that the ubiquitous, direct selection pressure to reduce the cost of connections between network nodes causes the emergence of modular networks. Computational evolution experiments with selection pressures to maximize network performance and minimize connection costs yield networks that are significantly more modular and more evolvable than control experiments that only select for performance. These results will catalyse research in numerous disciplines, such as neuroscience and genetics, and enhance our ability to harness evolution for engineering purposes.

## 1. Introduction

A long-standing, open question in biology is how populations are capable of rapidly adapting to novel environments, a trait called evolvability [1]. A major contributor to evolvability is the fact that many biological entities are modular, especially the many biological processes and structures that can be modelled as networks, such as metabolic pathways, gene regulation, protein interactions and animal brains [1–7]. Networks are modular if they contain highly connected clusters of nodes that are sparsely connected to nodes in other clusters [4,8,9]. Despite its importance and decades of research, there is no agreement on why modularity evolves [4,10,11]. Intuitively, modular systems seem more adaptable, a lesson well known to human engineers [12], because it is easier to rewire a modular network with functional subunits than an entangled, monolithic network [13,14]. However, because this evolvability only provides a selective advantage over the long term, such selection is at best indirect and may not be strong enough to explain the level of modularity in the natural world [4,10].

Modularity is probably caused by multiple forces acting to various degrees in different contexts [4], and a comprehensive understanding of the evolutionary origins of modularity involves identifying those multiple forces and their relative contributions. The leading hypothesis is that modularity mainly emerges because of rapidly changing environments that have common subproblems, but different overall problems [13,14]. Computational simulations demonstrate that in such environments (called modularly varying goals: MVG), networks evolve both modularity and evolvability [13,14]. By contrast, evolution in unchanging environments produces non-modular networks that are slower to adapt to new environments [13,14]. Follow-up studies support the modularity-generating force of MVG in nature: the modularity of bacterial metabolic networks is correlated with the frequency with which their environments change [15]. It is unknown how much natural modularity MVG can explain, however, because it is unclear how many biological environments change *modularly*, and whether they change at a high enough frequency for this force to play a significant role [11]. A related theory that also assumes a constantly changing environment and selection for evolvability is that modularity arises to enable modifying one subcomponent without affecting others [11]. There are other plausible hypotheses (reviewed in [4]), including that variation mechanisms, such as gene duplication, create a bias towards the generation of modular structures [4] and that modularity evolves because of selection to make phenotypes robust to environmental perturbations [10].

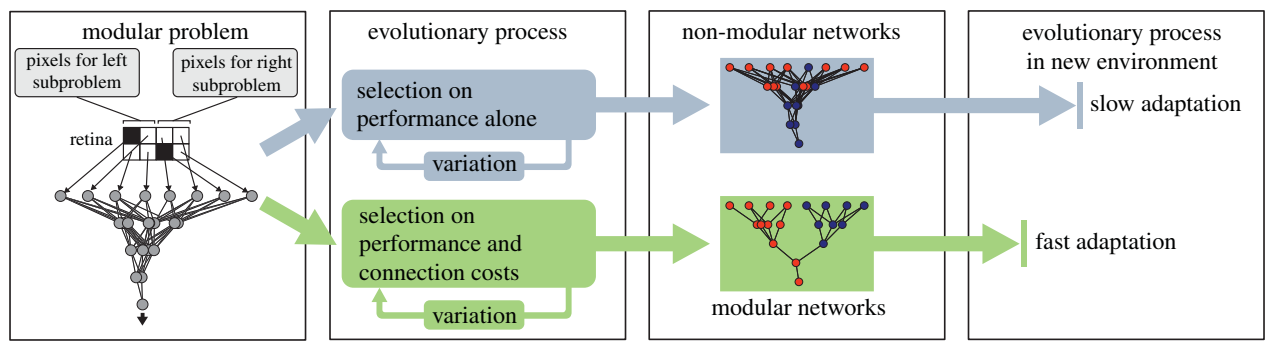

**Figure 1.** Main hypothesis. Evolving networks with selection for performance alone produces non-modular networks that are slow to adapt to new environments. Adding a selective pressure to minimize connection costs leads to the evolution of modular networks that quickly adapt to new environments.

We investigate an alternate hypothesis that has been suggested, but heretofore untested, which is that modularity evolves not because it conveys evolvability, but as a byproduct from selection to reduce connection costs in a network (figure 1) [9,16]. Such costs include manufacturing connections, maintaining them, the energy to transmit along them and signal delays, all of which increase as a function of connection length and number [9,17–19]. The concept of connection costs is straightforward in networks with physical connections (e.g. neural networks), but costs and physical limits on the number of possible connections may also tend to limit interactions in other types of networks such as genetic and metabolic pathways. For example, adding more connections in a signalling pathway might delay the time that it takes to output a critical response; adding regulation of a gene via more transcription factors may be difficult or impossible after a certain number of proximal DNA binding sites are occupied, and increases the time and material required for genome replication and regulation; and adding more protein–protein interactions to a system may become increasingly difficult as more of the remaining surface area is taken up by other binding interactions. Future work is needed to investigate these and other hypotheses regarding costs in cellular networks. The strongest evidence that biological networks face direct selection to minimize connection costs comes from the vascular system [20] and from nervous systems, including the brain, where multiple studies suggest that the summed length of the wiring diagram has been minimized, either by reducing long connections or by optimizing the placement of neurons [9,17–19,21–23]. Founding [16] and modern [9] neuroscientists have hypothesized that direct selection to minimize connection costs may, as a side-effect, cause modularity. This hypothesis has never been tested in the context of evolutionary biology. The most related study was on non-evolving, simulated neural networks with a specific within-life learning algorithm that produced more modularity when minimizing connection length in addition to performance [24], although the generality of the result was questioned when it was not replicated with other learning algorithms [25]. Without during-life learning algorithms, carefully constructed MVG environments or mutation operators strongly biased towards creating modules, attempts to evolve modularity in neural networks have failed [10,26,27].

Given the impracticality of observing modularity evolve in biological systems, we follow most research on the subject by conducting experiments in computational systems with evolutionary dynamics [4,11,13]. Specifically, we use a well-studied system from the MVG investigations [13,14,27]: evolving networks to solve pattern-recognition tasks and Boolean logic tasks (§4). These networks have inputs that sense the environment and produce outputs (e.g. activating genes, muscle commands, etc.), which determine performance on environmental problems. We compare a treatment where the fitness of networks is based on performance alone (PA) to one based on two objectives: maximizing performance and minimizing connection costs (P&CC). A multi-objective evolutionary algorithm is used [28] with one (PA) or two (P&CC) objectives: to reflect that selection is stronger on network performance than connection costs, the P&CC cost objective affects selection probabilistically only 25 per cent of the time, although the results are robust to substantial changes to this value (§4). Two example connection cost functions are investigated. The default one is the summed squared length of all connections, assuming nodes are optimally located to minimize this sum (§4), as has been found for animal nervous systems [17,18,29,30]. A second measure of costs as solely the number of connections yields qualitatively similar results to the default cost function, and may better represent biological networks without connections of different lengths. More fit networks tend to have more offspring (copies that are randomly mutated), and the cycle repeats for a preset number of generations (figure 1, §4). Such computational evolving systems have substantially improved our understanding of natural evolutionary dynamics [4,11,13,14,31,32].

The main experimental problem involves a network that receives stimuli from eight inputs [13]. It can be thought of as an eight-pixel retina receiving visual stimuli, although other analogies are valid (§4), such as a genetic regulatory network exposed to environmental stimuli. Patterns shown on the retina's left and right halves may each contain an 'object', meaning a pattern of interest (figure 2a). Networks evolve to answer whether an object is present on both the left *and* right sides of the retina (the L-AND-R environment) or whether an object is displayed on *either* side (the L-OR-R environment). Which patterns count as an object on the left and right halves are slightly different

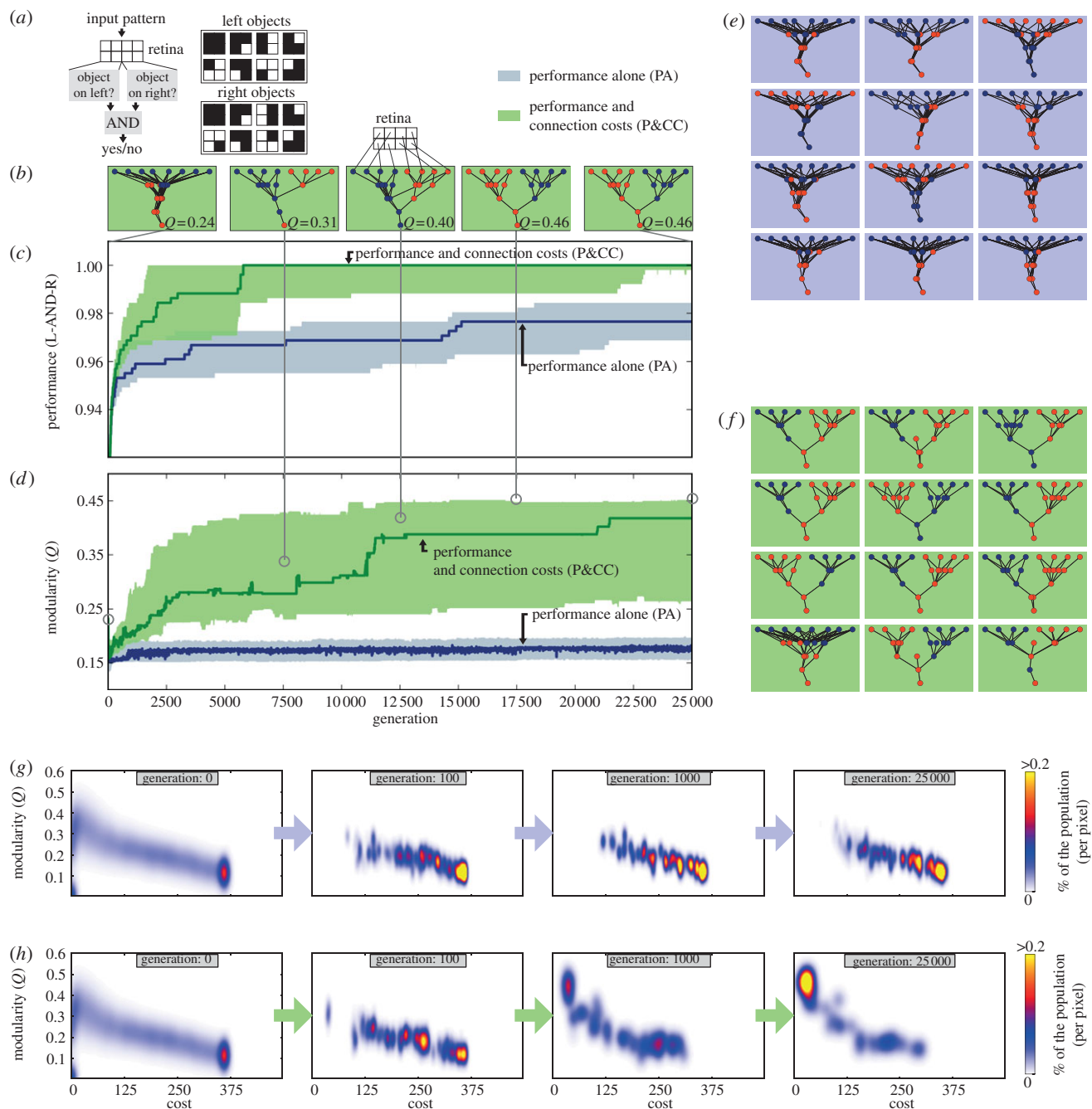

**Figure 2.** The addition of connection costs leads to higher-performing, functionally modular networks. (a) Networks evolve to recognize patterns (objects) in an eight-pixel retina. The problem is modularly decomposable because whether an object exists on the left and right sides can be separately determined before combining that information to answer whether objects exist on *both* sides (denoted by the AND logic function). (b) Networks from an example trial become more modular across evolutionary time (also see the electronic supplementary material, video S1) with a pressure to minimize connection costs in addition to performance (P&CC). (c) Median performance ($\pm 95\%$ bootstrapped confidence intervals) per generation of the highest-performing network of each trial, which is perfect only when minimizing connection costs in addition to performance. (d) Network modularity, which is significantly higher in P&CC trials than when selecting for performance alone (PA). (e) The 12 highest-performing PA networks, each from a separate trial. (f) The 12 highest-performing P&CC networks, which are functionally modular in that they have separate modules for the left and right subproblems. Nodes are coloured according to membership in separate partitions when making the most modular split of the network (see text). The final networks of all 50 trials are visualized in the electronic supplementary material, figure S1. (g,h) Cost and modularity of PA and P&CC populations across generations, pooled from all 50 trials. A connection cost pushes populations out of high-cost, low-modularity regions towards low-cost, modular regions. Figure 3 shows the fitness potential of each map area.

(figure 2a). Each network iteratively sees all possible 256 input patterns and answers true ($\geq 0$) or false ($< 0$). Its performance is the percentage of correct answers, which depends on which nodes are connected, how strongly, and whether those connections are inhibitory or excitatory (§4). Networks are randomly generated to start each experiment. Their connections stochastically mutate during replication (§4). Network modularity is evaluated with an efficient approximation [33,34] of the widely used modularity metric $Q$, which first optimally divides networks into modules then measures the difference between the number of edges within each module and the number expected for random networks with the same number of edges [33,34].

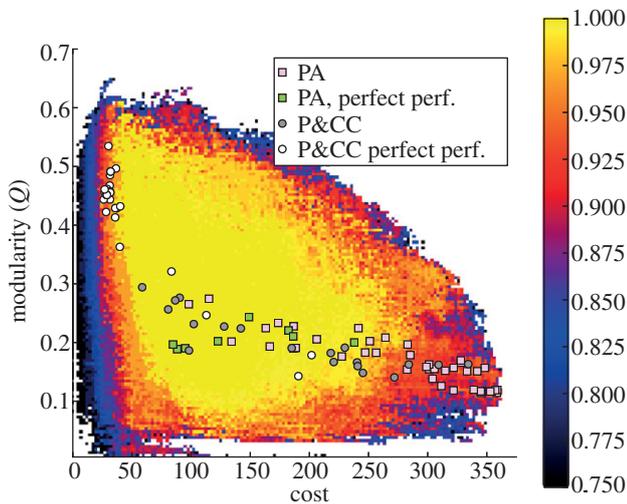

**Figure 3.** The highest-performing networks found for each combination of modularity and cost for the retina problem. Colours indicate the highest-performing network found at that point in the modularity versus cost space, with yellow representing perfect performance. This map has been generated using the MOLE algorithm (§4). The best-performing network at the end of each of the 50 PA and P&CC runs are overlaid on the map. Networks with perfect performance exist throughout the space, which helps explain why modularity does not evolve when there is selection based on performance alone. Below a cost threshold of around 125 there is an inverse correlation between cost and modularity for perfectly performing networks. The lowest cost networks—those with the shortest summed lengths—that are high-performing are modular.

## 2. Results

After 25 000 generations in an unchanging environment (L-AND-R), treatments selected to maximize performance and minimize connection costs (P&CC) produce significantly more modular networks than treatments maximizing performance alone (PA) (figure 2d, $Q = 0.42$, 95% CI [0.25,0.45] vs. $Q = 0.18$[0.16, 0.19], $p = 8 \times 10^{-09}$ using Matlab's Mann–Whitney–Wilcoxon rank sum test, which is the default statistical test unless otherwise specified). To test whether evolved networks exhibit *functional modularity* corresponding to the left–right decomposition of the task we divide networks into two modules by selecting the division that maximizes $Q$ and colour nodes in each partition differently. Left–right decomposition is visually apparent in most P&CC trials and absent in PA trials (figure 2e,f). Functional modularity can be quantified by identifying whether left and right inputs are in different partitions, which occurs in 56 per cent of P&CC trials and never with PA (Fisher's exact test, $p = 4 \times 10^{-11}$). Pairs of perfect sub-solution nodes—whose outputs perfectly answer the left and right subproblems—occur in 39 per cent of P&CC trials and 0 per cent of PA trials (Fisher's exact test, $p = 3 \times 10^{-6}$, electronic supplementary material, figure S1).

Despite the additional constraint, P&CC networks are significantly higher-performing than PA networks (figure 2c, electronic supplementary material, figure S13). The median-performing P&CC network performs perfectly (1.0[1.0, 1.0]), but the median PA network does not (0.98[0.97, 0.98], $p = 2 \times 10^{-05}$). P&CC performance may be higher because its networks have fewer nodes and connections (see the electronic supplementary material, figure S8b,c), meaning fewer parameters to optimize. Modular structures are also easier to adapt since mutational effects are smaller, being confined to subcomponents [8]. While it is thought that optimal, non-modular solutions usually outperform optimal, modular designs, such 'modularity overhead' only exists when comparing *optimal* designs, and is not at odds with the finding that *adaptation* can be faster and ultimately more successful with a bias towards modular solutions [8].

To better understand why the presence of a connection cost increases performance and modularity, we searched for the highest-performing networks at all possible combinations of modularity and cost (§4). For high-performing networks, there is an inverse correlation between cost and modularity, such that the lowest-cost networks are highly modular (figure 3). Many runs in the P&CC treatment evolved networks in this region whereas the PA treatments never did. There are also many non-modular, high-cost networks that are high-performing, helping one explain why modularity does not evolve due to performance alone (figure 3). Comparing PA versus P&CC populations across generations reveals that a connection cost pushes populations out of high-cost, low-modularity areas of the search space into low-cost, modular areas (figure 2g,h). Without the pressure to leave high-cost, low-modularity regions, many PA networks remain in areas that ultimately do not contain the highest-performing solutions (figure 3, pink squares in the bottom right), further explaining why P&CC treatments have higher performance. We also found evidence of an inverse correlation between the total cost of a network and modularity in randomly generated networks, irrespective of performance, supporting the intuition that low-cost networks are more likely to be modular (see the electronic supplementary material, figure S12).

P&CC networks are also more evolvable than PA networks. We ran additional trials until 50 P&CC and 50 PA trials each had a perfectly performing network (§4) and transferred these networks into the L-OR-R environment, which has the same subproblems in a different combination (see the electronic supplementary material, figure S6). The presence (P&CC) or absence (PA) of a connection cost remained after the environmental change. We performed 50 replicate experiments for each transferred network. We also repeated the experiment, except first evolving in L-OR-R and then transferring to L-AND-R. In both experiments, P&CC networks exhibit greater evolvability than PA by requiring fewer generations to adapt to the new environment (figure 4a, L-AND-R → L-OR-R: 3.0[2.0, 5.0] versus 65[62, 69], $p = 3 \times 10^{-78}$; L-OR-R → L-AND-R: 12.0[7.0, 21.0] versus 222.5[175.0, 290.0], $p = 9 \times 10^{-120}$). Modular networks thus evolve because their sparse connectivity has lower connection costs, but such modularity also aids performance and evolvability because the problem is modular.

Minimizing connection costs can work in conjunction with other forces to increase modularity. Modularity levels are higher when combining P&CC with MVG environments (figure 4b: solid versus dotted green line, $p = 3 \times 10^{-5}$). Overall, P&CC (with or without MVG) yields similar levels of modularity as MVG at its strongest, and significantly more when rates of environmental change are too slow for the MVG effect to be strong (figure 4: green lines versus blue solid line).

P&CC modularity is also higher than PA even on problems that are non-modular (figure 5a, $p = 5.4 \times 10^{-18}$). As to be expected, such modularity is lower than on modular problems ($p = 0.0011$, modular retina versus non-modular retina). This non-modular problem involves answering whether any four pixels were on (black), which is non-modular because it requires information from all retina inputs. As mentioned previously, performance and modularity are also significantly higher with an alternate connection cost function based on the number of connections (P&CC-NC) instead of the length of connections (figure 6). We also verified that modularity and performance are not higher simply because a second objective is used (figure 6). We further tested whether modularity arises even when the inputs for different modules are not geometrically separated, which is relevant when cost is a function of connection length. Even in experiments with randomized input coordinates (§4), a connection cost significantly increased performance (1.0[0.98, 1.0], $p = 0.0012$) and modularity ($Q = 0.35[0.34, 0.38]$, $p = 1 \times 10^{-9}$), and performance and modularity scores were not significantly different than P&CC without randomized coordinates (see the electronic supplementary material, figure S7).

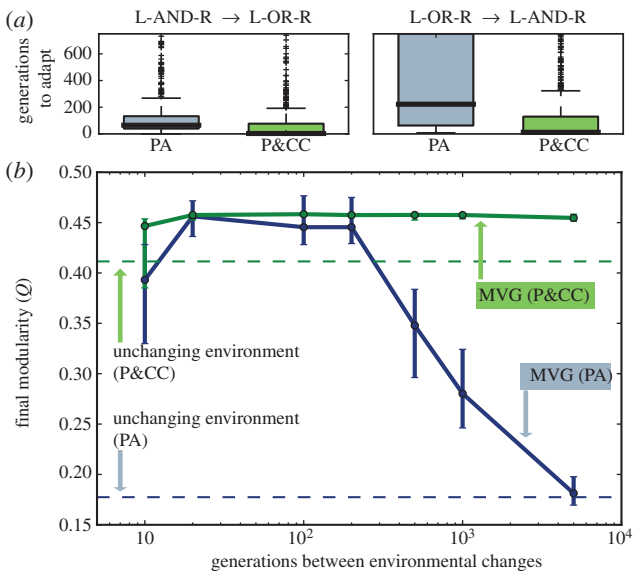

**Figure 4.** Evolving with connection costs produces networks that are more evolvable. (a) P&CC networks adapt faster to new environments than PA networks. Organisms were first evolved in one environment (e.g. L-AND-R) until they reached perfect performance and then transferred to a second environment (e.g. L-OR-R). Thick lines are medians, boxes extend from 25th to 75th data percentiles, thin lines mark 1.5 × IQR (interquartile range), and plus signs represent outliers. Electronic supplementary material figure S6 is a zoomed-out version showing all of the data. (b) P&CC networks in an unchanging environment (dotted green line) have similar levels of modularity to the highest levels produced by MVG (solid blue line). Combining MVG with P&CC results in even higher modularity levels (solid green line), showing that the forces combined are stronger than either alone.

All the results presented so far are qualitatively similar in a different model system: evolving networks to solve Boolean logic tasks. We tested two fully separable problems: one with five 'exclusive or' (XOR) logic modules (figure 5b), and another with hierarchically nested XOR problems (figure 5c). P&CC created separate modules for the decomposed problems in nearly every trial, whereas PA almost never did (see the electronic supplementary material, figures S2 and S3). P&CC performance was also significantly higher (figure 5b,c), and there was an inverse correlation between cost and modularity (see the electronic supplementary material, figure S10). After reading a preprint of this manuscript, a different research group replicated the main result in a different domain: they found that a connection cost causes modularity to evolve when optimizing computer chip architectures [35]. Confirming the generality of the finding that connection costs improve adaptation rates and that high-performing, low-cost networks are modular is an interesting area for future research.

## 3. Discussion and conclusion

Overall, this paper supports the hypothesis that selection to reduce connection costs causes modularity, even in unchanging environments. The results also open new areas of research into identifying connection costs in networks without physical connections (e.g. genetic regulatory networks) and investigating whether pressures to minimize connection costs may explain modularity in human-created networks (e.g. communication and social networks).

It is tempting to consider any component of modularity that arises due to minimizing connection costs as a 'spandrel', in that it emerges as a byproduct of selection for another trait [36,37]. However, because the resultant modularity produces evolvability, minimizing connection costs may serve as a bootstrapping process that creates initial modularity that can then be further elevated by selection for evolvability. Such hypotheses for how modularity initially arises are needed, because selection for evolvability cannot act until enough modularity exists to increase the speed of adaptation [4].

Knowing that selection to reduce connection costs produces modular networks will substantially advance fields that harness evolution for engineering, because a longstanding challenge therein has been evolving modular designs [8,10,27,38]. It will additionally aid attempts to evolve accurate models of biological networks, which catalyse medical and biological research [2,9,39]. The functional modularity generated also makes synthetically evolved networks easier to understand. These results will thus generate immediate benefits in many fields of applied engineering, in addition to furthering our quest to explain one of nature's predominant organizing principles.

## 4. Methods

### (a) Experimental parameters
Each treatment is repeated 50 times with different stochastic events (i.e. different random number generator seeds). Analyses and visualizations are of the highest-performing network per trial with ties broken randomly. The main experiments (retina, non-modular retina, 5-XOR, and hierarchical XOR) last 25 000 generations and have a population size of 1000.

### (b) Statistics
For each statistic, we report the median ±95% bootstrapped confidence intervals of the median (calculated by resampling the data 5000 times). In plots, these confidence intervals are smoothed with a median filter (window size = 200) to remove sampling noise. Statistical significance is assessed using Matlab's Mann–Whitney–Wilcoxon rank sum test.

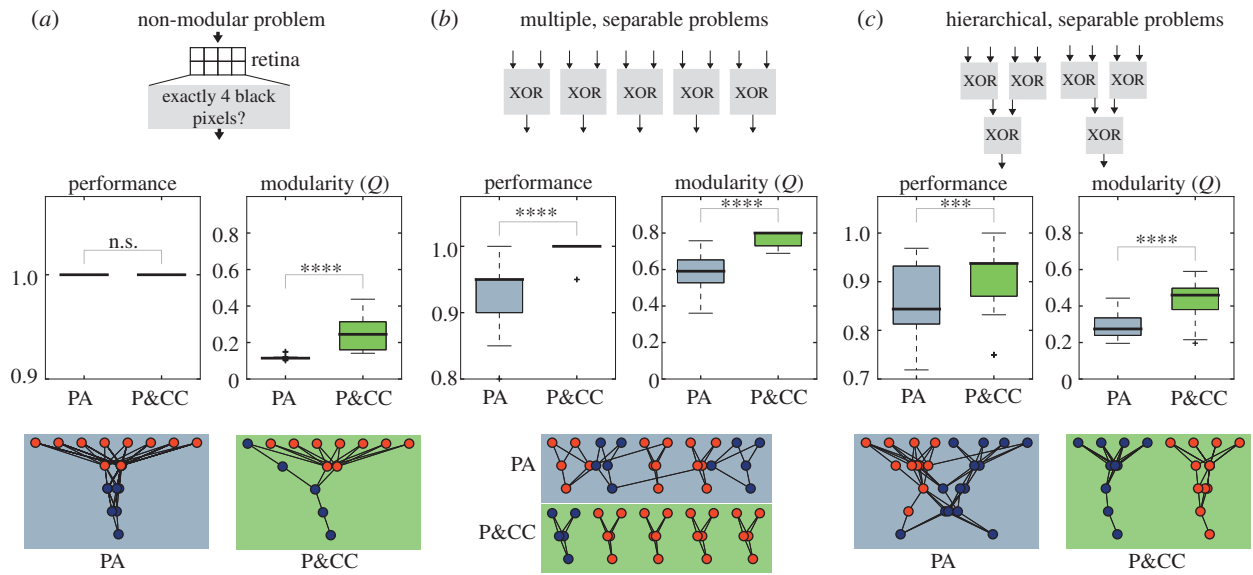

**Figure 5.** Results from tests with different environmental problems. (*a*) Even on a non-modular problem, modularity is higher with P&CC, though it is lower than for modular problems. (*b*,*c*) P&CC performs better, is more modular, and has better functional decomposition than PA when evolving networks to solve five separate XOR functions and hierarchically nested XOR functions. The examples are the final, highest-performing networks per treatment. Electronic supplementary material figures S2–S4 show networks from all trials. Three and four asterisks indicate $p$ values less than 0.001 and 0.0001, respectively, and n.s. indicates no significant difference.

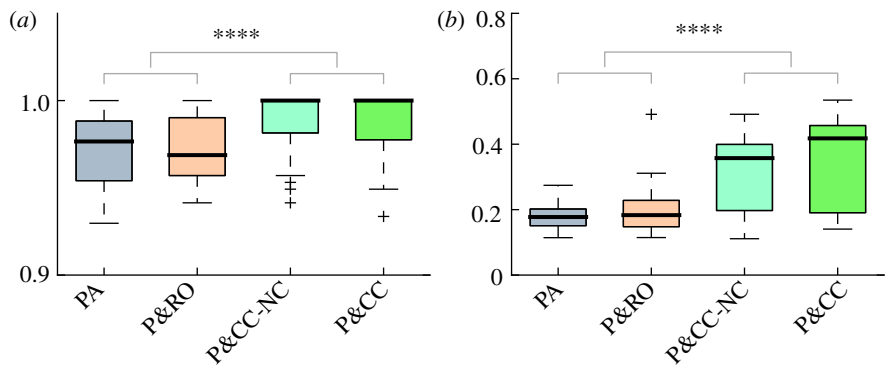

**Figure 6.** Alternate cost functions. Performance (*a*) and modularity (*b*) are significantly higher ($p < 0.0001$) either with a cost function based on the length (P&CC) or number (P&CC-NC) of connections versus performance alone (PA) or performance and a random objective (P&RO). P&RO assigned a random number to each organism instead of a connection cost score and maximized that random number. Electronic supplementary material figure S7 contains visualizations of all P&CC-NC networks.

### (c) Length cost

For P&CC treatments, prior to calculating connection costs, we place nodes in positions optimal for minimizing connection costs given the topology of the network, which is biologically motivated [17,18,29] and can be solved for mathematically [29]. Inputs and outputs are at fixed locations (see §4*d*). Visualizations reflect these node placements.

### (d) Geometric coordinates

Nodes exist at two-dimensional (i.e. $x,y$) Cartesian locations. The geometric coordinates of the inputs and outputs for all problems were fixed throughout evolution, including the treatment where the within-row location of inputs are randomized at the beginning of each separate trial. The inputs for all problems have $y$-values of 0. For the retina problem, the $x$-values for the inputs are $\{-3.5, -2.5, \ldots, 3.5\}$ and the output is at $\{4, 0\}$. For the problem with 5 XOR modules, the $x$-values for the inputs are $\{-4.5, -3.5, \ldots, 4.5\}$ and the outputs all have $y$-values of 2 with $x$-values of $\{-4, -2, 0, 2, 4\}$. For the problem with decomposable, hierarchically nested XOR functions, the $x$-values for the inputs are $\{-3.5,$ $-2.5, \ldots, 3.5\}$ and the outputs all have $y$-values of 4 with $x$-values of $\{-2, 2\}$. The geometric location of nodes is consequential only when there is a cost for longer connections (i.e. the main P&CC treatment).

### (e) Evolvability experiment

The evolvability experiments (figure 4*a*) are described in electronic supplementary material, figure S6. To obtain 50 trials that each had a perfectly performing network in L-AND-R and L-OR-R, respectively, took 110 and 116 trials for P&CC and 320 and 364 trials for PA. One thousand clones of each of these networks then evolved in the alternate environment until performance was perfect or 5000 generations passed.

### (f) Biological relevance of network models

This section provides a brief overview of the network models in this paper. A more complete review of network models is provided in [2,40–42] and the references therein.

Network models can represent many types of biological processes by representing interactions between components

[2,41,42]. Examples of biological systems that are commonly modelled as networks are genetic, metabolic, protein interaction and neural networks. All such networks can be represented abstractly as nodes representing components, such as neurons or genes, and the interactions between such components, such as a gene inhibiting another gene. The weight of connections indicates the type and strength of interactions, with positive values indicating activation, negative values indicating inhibition, and the magnitude of the value representing the strength of the interaction.

Multiple nodes can connect to form a network (e.g. electronic supplementary material, figure S11b). Typically, information flows into the network via *input* nodes, passes through *hidden* nodes, and exits via *output* nodes. Examples include a gene regulatory network responding to changing levels of environmental chemicals or a neural network responding to visual inputs from the retina and outputting muscle commands.

### (g) Network model details

Our model of a network is a standard, basic one used in machine learning [40], systems biology [2,13,14,43] and computational neuroscience [44]. It has also been used in previous landmark studies on the evolution of modularity [13,11]. The networks are *feed-forward*, meaning that nodes are arranged into layers, such that a node in layer $n$ receives incoming connections only from nodes in layer $n-1$ and has outgoing connections only to nodes in layer $n+1$ (see the electronic supplementary material, figure S11b). The maximum number of nodes per hidden layer is 8/4/2 for the three hidden layers in the retina problem, 8 for the single hidden layer in the 5-XOR problem, and 8/4/4 for the three hidden layers in the hierarchical XOR problem. The possible values for connection weights are the integers $-2, -1, 1$ and $2$. The possible values for thresholds (also called biases) are the integers $-2, -1, 0, 1$ and $2$. Information flows through the network in discrete time steps one layer at a time. The output of each node in the network is the following function of its inputs: $y_j = \tanh(\lambda(\sum_{i \in I_j} w_{ij} y_i + b))$ where $y_j$ is the output of node $j$, $I$ is the set of nodes connected to $j$, $w_{ij}$ is the strength of the connection between node $i$ and node $j$, $y_i$ is the output of node $i$, and $b$ is a threshold (also called a bias) that determines at which input value the output transitions from negative to positive. The $\tanh(x)$ transfer function ensures an output range of $[-1, 1]$. $\lambda$ (here, 20) determines the slope of the transition between these inhibitory and excitatory extremes (see the electronic supplementary material, figure S11c). This network model can approximate any function with an arbitrary precision provided that it contains enough hidden nodes [45].

### (h) Evolutionary algorithm

The evolutionary algorithm is based on research into algorithms inspired by evolution that simultaneously optimize several objectives, called *multi-objective algorithms* [28]. These algorithms search for, but are not guaranteed to find, the set of optimal *trade-offs*: i.e. solutions that cannot be improved with respect to one objective without decreasing their score with respect to another one. Such solutions are said to be on the *Pareto Front* [28], described formally below. These algorithms are more general than algorithms that combine multiple objectives into a single, weighted fitness function, because the latter necessarily select one set of weights for each objective, whereas multi-objective algorithms explore all possible trade-offs between objectives [28].

The specific multi-objective algorithm in this paper is the widely used Non-dominated Sorting Genetic Algorithm, version II (NSGA-II) [28] (see the electronic supplementary material, figure S11a). As with most modern multi-objective evolutionary algorithms, it relies on the concept of Pareto dominance, defined as follows.

An individual $x^*$ is said to dominate another individual $x$, if both conditions 1 and 2 are true: (1) $x^*$ is not worse than $x$ with respect to any objective; (2) $x^*$ is strictly better than $x$ with respect to at least one objective.

However, this definition puts the same emphasis on all objectives. In the present study, we take into account that the first objective (performance) is more important than the second objective (optimizing connection cost). To reflect this, we use a stochastic version of Pareto dominance in which the second objective is only taken into account with a given probability $p$. Lower values of $p$ cause lower selection pressure on the second objective. Our results are robust to alternate values of $p$, including up to $p = 1.0$ for static environments (see the electronic supplementary material, figure S5a–d) and $p = 0.95$ for environments with modularly varying goals (see the electronic supplementary material, figure S5e).

This stochastic application of the second objective is implemented as follows. Let $r$ denote a random number in $[0; 1]$ and $p$ the probability to take the second objective into account. A solution $x^*$ is said to *stochastically* dominate another solution $x$, if one of the two following conditions is true: (1) $r > p$ and $x^*$ is better than $x$ with respect to the first objective; (2) $r \leq p$ and $x^*$ is not worse than $x$ with respect to either objective and $x^*$ is better than $x$ with respect to at least one objective.

Stochastic Pareto dominance is used in the algorithm twice (see the electronic supplementary material, figure S11a). (i) To select a parent for the next generation, two individuals $x_1$ and $x_2$ are randomly chosen from the current population; if $x_1$ stochastically dominates $x_2$, then $x_1$ is selected, if $x_2$ stochastically dominates $x_1$, then $x_2$ is selected. If neither dominates the other, the individual selected is the one which is in the less crowded part of the objective space [28]. (ii) To rank individuals, the set of stochastically non-dominated solutions is first identified and called the first Pareto layer (rank = 1, e.g. $l_1$ in electronic supplementary material, figure S11a); these individuals are then removed and the same operation is performed to identify the subsequent layers (additional ranks corresponding to $l_2$, $l_3$, etc. in electronic supplementary material, figure S11a).

### (i) Mutational effects

Mutations operate in essentially the same way as in the study by Kashtan & Alon [13]. In each generation, every new network is randomly mutated (see the electronic supplementary material, figure S11a). Four kinds of mutation are possible, which are not mutually exclusive: (i) each network has a 20 per cent chance of having a single connection added. Connections are added between two randomly chosen nodes that are not already connected and belong to two consecutive layers (to maintain the feed-forward property described previously); (ii) each network has a 20 per cent chance of a single, randomly chosen connection being removed; (iii) each node in the network has a $1/24 = 4.16$ per cent chance of having its threshold (also called its bias) incremented or decremented, with both options equally probable; five values are available $\{-2, -1, 0, 1, 2\}$; mutations that produce values higher or lower than these five values are ignored; (iv) each connection in the network has a separate probability of being incremented or decremented of $2.0/n$, where $n$ is the total number of connections of the network. Four values are available $\{-2, -1, 1, 2\}$; mutations that produce values higher or lower than these four values are ignored.

The results in this manuscript are robust to varying these parameters. Because having more mutational events that remove connections than add them might also produce sparsely connected, modular networks, we repeated the main experiment with mutation rates biased to varying degrees (see the electronic supplementary material, figure S9). These experiments show that even having remove–connection events be an order of magnitude more probable than add-connection events does not reduce the number of connections or produce modular networks. In each of the experiments with biased mutation rates, the modularity and performance of the P&CC treatment with default mutation values was significantly greater than the PA treatment with biased mutation rate values.

Our main results are qualitatively the same when weights and biases are real numbers (instead of integers) and mutated via Gaussian perturbation. Nodes are never added nor removed. For clarity, following [13], nodes without any connections are not displayed or included in results.

### (j) Multi-objective landscape exploration algorithm

To better understand evolving systems it would be helpful to visualize important constraints, trade-offs, and other correlations between different phenotypic dimensions in evolving populations (e.g. in this study, the performance, modularity, and cost for each possible network). If the search space is small enough, such values can be determined by exhaustively checking every possible solution. Such an approach is intractable for the problems in this manuscript. For example, for the main problem in the paper, which is the retina problem, the number of possible weights is $(8 \times 8) + (8 \times 4) + (4 \times 2) + (2 \times 1) + 23 = 129$, owing to the number of nodes in the input, hidden, and output layers, as well as the bias for each of the 23 possible nodes. Each of these weights can be one of four values or a zero if no connection exists, and biases can be one of five values, leading to a search space of $5^{129} = 10^{90}$. Given that it takes on average 0.0013 s to assess the fitness of a solution across all possible 256 inputs using a modern computer, it would take $4.1 \times 10^{79}$ years of computing time to exhaustively determine the performance, modularity and cost for each solution in the search space.

Because it is infeasible to exhaustively search the space, we randomly sampled it to see whether we would find high-performing solutions with a variety of cost and modularity scores. Specifically, we randomly generated more than two billion solutions, but every solution had poor performance. The highest-performing solution gave the correct answer for only 62 out of 256 retina patterns (24.2%), which is far below the performance of 93 per cent or greater for solutions routinely discovered by the evolutionary algorithm (see the electronic supplementary material, figure S1). We thus concluded that randomly sampling the space would not lead to the discovery of high-performing solutions.

We therefore designed an algorithm to find high-performing solutions with different combinations of modularity and length scores. We call this algorithm Multi-Objective Landscape Exploration (MOLE). It is a multi-objective optimization search [28] (see the electronic supplementary material, figure S11a) with two objectives. The first objective prioritizes individuals that have high performance. The second objective prioritizes individuals that are far away from other individuals already discovered by the algorithm, where distance is measured in a Cartesian space with connection costs on the $x$-axis and modularity on the $y$-axis. Algorithms of this type have been shown to better explore the search space because they are less susceptible to getting stuck in local optima [46]. Thus, unlike a traditional evolutionary algorithm that will only be drawn to a type of solution if there is a fitness gradient towards that type of solution, MOLE searches for high-performing solutions for every possible combination of modularity and cost scores. While this algorithm is not guaranteed to find the optimal solution at each point in the space, it provides a focused statistical sampling of how probable it is to discover a high-quality solution in each area of the search space. The MOLE maps in this paper (figure 3 and electronic supplementary material, figure S10) show the highest-performing network at each point in this Cartesian space found in 30 separate runs.

### (k) Video of networks from each treatment evolving across generations

A video is provided to illustrate the change in networks across evolutionary time for both the PA and P&CC treatments. In that the networks are visualized as described in the text. The video can be downloaded at: http://dx.doi.org/10.5061/dryad.9tb07.

### (l) Experimental data and source code

All of the experimental data, source code and analysis scripts are freely available in a permanent online archive at http://dx.doi.org/10.5061/dryad.9tb07.

Funding provided by an NSF Postdoctoral Research Fellowship in Biology to J.C. (DBI-1003220) and NSF CDI Grant ECCS 0941561. J.B.M. is supported by the ANR (project Creadapt ANR-12-JS03-0009). Thanks are due to Shimon Whiteson, Rafael Sanjuán, Stéphane Doncieux, Dusan Misevic, Jeff Barrick, Charles Ofria and the reviewers for helpful comments.